\documentclass[sn-mathphys,Numbered]{sn-jnl}

\usepackage{graphicx}%
\usepackage{multirow}%
\usepackage{amsmath,amssymb,amsfonts}%
\usepackage{amsthm}%
\usepackage{mathrsfs}%
\usepackage[title]{appendix}%
\usepackage{xcolor}%
\usepackage{textcomp}%
\usepackage{manyfoot}%
\usepackage{booktabs}%
\usepackage{algorithm}%
\usepackage{algorithmicx}%
\usepackage{algpseudocode}%
\usepackage{listings}%

\theoremstyle{thmstyleone}%
\theoremstyle{thmstyletwo}%

\theoremstyle{thmstylethree}%

\begin{document}

\title[Article Title]{Cross Modality Medical Image Synthesis for Improving Liver Segmentation}



\author[1]{\fnm{Muhammad} \sur{Rafiq}}\email{m.rafiq1882@gmail.com}

\author*[2]{\fnm{Hazrat} \sur{Ali}}\email{hazrat.ali@live.com, ali.hazrat@stir.ac.uk}

\author[1]{\fnm{Ghulam} \sur{Mujtaba}}\email{gmujtaba@cuiatd.edu.pk}
\author[3]{\fnm{Zubair} \sur{Shah}}\email{zshah@hbku.edu.qa}

\author[1]{\fnm{Shoaib} \sur{Azmat}}\email{shoaibazmat@cuiatd.edu.pk}

\affil[1]{\orgdiv{Department of Electrical and Computer Engineering}, \orgname{COMSATS University Islamabad}, \orgaddress{\street{Abbottabad Campus}, \city{Abbottabad}, \country{Pakistan}}}

\affil[2]{\orgdiv{Computing Science and Mathematics, University of Stirling}, \orgaddress{\city{Stirling}, \country{United Kingdom}}}

\affil[3]{\orgdiv{College of Science and Engineering}, \orgname{Hamad Bin Khalifa University}, \orgaddress{\street{Education City}, \city{Doha}, \country{Qatar}}}



\abstract{Deep learning-based computer-aided diagnosis (CAD) of medical images requires large datasets. However, the lack of large publicly available labeled datasets limits the development of deep learning-based CAD systems. Generative Adversarial Networks (GANs), in particular, CycleGAN, can be used to generate new cross-domain images without paired training data. However, most CycleGAN-based synthesis methods lack the potential to overcome alignment and asymmetry between the input and generated data. We propose a two-stage technique for the synthesis of abdominal MRI using cross-modality translation of abdominal CT. We show that the synthetic data can help improve the performance of the liver segmentation network. We increase the number of abdominal MRI images through cross-modality image transformation of unpaired CT images using a CycleGAN inspired deformation invariant network called EssNet. Subsequently, we combine the synthetic MRI images with the original MRI images and use them to improve the accuracy of the U-Net on a liver segmentation task. We train the U-Net on real MRI images and then on real and synthetic MRI images. Consequently, by comparing both scenarios, we achieve an improvement in the performance of U-Net. In summary, the improvement achieved in the Intersection over Union (IoU) is 1.17\%. The results show potential to address the data scarcity challenge in medical imaging.}




\maketitle

\section{Introduction}\label{sec1}
Deep learning models have advanced the current state of artificial intelligence (AI) in medical imaging. In general, we require a significant number of labeled images to train a deep learning model. However, acquiring medical image data with ground-truth labels is costly and time-consuming, as one must rely on manual annotation by trained radiologists. In addition, privacy concerns also limit the public sharing of medical data. On the other hand, training a deep learning model on a small number of images limits the generalization of the model. For small medical datasets, synthesizing newer images using the available cross-modality data may help to improve the model’s training. Furthermore, certain features of human tissues may be better visualized using only certain modalities. For example, compared to computed tomography (CT) scans, magnetic resonance imaging (MRI) can provide better insights into soft tissues \cite{1}. However, an MRI scanner is a complex device that is costly to operate and only available at specialized centers \cite{2}. On the other hand, CT is comparatively less expensive than MRI and is, therefore, more accessible \cite{3}. One possible approach to increasing the number of MRI images is to transform CT into MRI. This domain-to-domain translation can aid in gaining additional MRI data and hence, reveal features in human tissues that would otherwise be not available. Consequently, this increase in data can improve the performance of a deep learning model. It is important to note that increasing images by this domain-to-domain image synthesis is different from data augmentation; in data augmentation, the changes are in the geometric and photometric transformations of the same image, while in image synthesis, new images are generated from original images to increase the number of images in a dataset. 

Recently Generative adversarial networks (GANs) have demonstrated good potential to generate synthetic image data (e.g., \cite{4}, \cite{5}, \cite{6}, \cite{7}). GAN consists of a generative network G and a discriminative network D. The generative network G takes the noise data and produces synthetic data, and the discriminative network D takes input both from the generator G and from the training dataset and outputs the probability of whether the input has come from the real data \cite{8}. In each iteration of the training, the discriminator is updated to make it better at discriminating between synthetic and real samples, and accordingly, the generator is updated based on its performance in fooling the discriminator. In other words, the two models are trained together in a two-player minimax game. The training of GAN \cite{8} is prone to instability. DCGAN \cite{9} introduced a set of constraints to address the instability problem. The GAN also has the problem of generating images unconditionally, i.e., with no control over the output. The CGAN model provided control over the data synthesis process by conditioning the GAN based on supplementary information \cite{10}. This conditional supplementary information may be a class label or data from another modality. ACGAN \cite{11} takes both the class label and the noise as input and generates output images. The above conditional GANs are application specific. To generalize CGAN, image-to-image translation GAN (pix2pix GAN) maps an input image from one domain to an output image in a different domain. The pix2pix GAN has demonstrated effectiveness on a wide variety of problems \cite{12}. The pix2pix-GAN was recently used by \cite{13} to improve the registration and segmentation of MRI images. Pix2pix-GAN can be utilized to convert a T1-weighted MRI image into a T2-weighted MRI image, as demonstrated by \cite{14}. {Similarly, Liu et al. \cite{31} synthesized brain MRI image types, including T1, T2, DWI, and Flair, using paired images based on a pix2pix-GAN-inspired architecture.} Other architectures inspired by pix2pix-GAN include MI-GAN \cite{15}, which generates retinal vessel images using paired training data to improve the performance of the vessel segmentation, while CovidGAN \cite{16} generates chest X-ray (CXR) images using paired data to improve the CNN performance for COVID-19 detection. {Furthermore, in \cite{35} and \cite{36}, paired CT and MRI images are used to improve the performance of multiorgan pelvic segmentation.} However, the pix2pix-GAN requires paired samples of data from the source and target domains, which are not commonly available in the medical imaging field.

The CycleGAN model extracts features from the input domain image and learns to translate these features to the target domain image without any paired training samples \cite{17}. Shen et al., \cite{18} used CycleGAN as the baseline framework for MRI image-to-image translation (T1 to T2 and vice versa). Zhang et al., \cite{19} used a large set of publicly available pediatric structural brain MRI images to construct a switchable CycleGAN model for image synthesis between multi-contrast brain MRI (T1w and T2w) images. SLA-StyleGAN \cite{20} used unpaired data to synthesize skin lesion images of high quality, which in turn improved the classification accuracy of skin lesions. {Park et al. \cite{30} proposed an improved variant of CycleGAN based on contrastive learning. Guo et al. \cite{34} used the model from Park et al., \cite{30} for data augmentation to improve the accuracy of kidney segmentation. They generated more kidney ultrasound images using unpaired CT images to improve segmentation accuracy on U-Net using both real and synthesized ultrasound images.
Hong et al. \cite{32} developed a specialized domain adaptation framework based on adversarial learning between unpaired CT and MRI. They synthesized MRI images from CT and segmented the liver in abdominal MRI using only CT labels, without any MRI labels.}

In short, for synthesizing cross-domain medical images (e.g., CT to MRI, MRI to CT, CT to PET, etc.), CycleGAN and its variants mentioned above have proven to be effective for domain-to-domain translation tasks on unpaired data. However, most CycleGAN based synthesis methods lack the potential to overcome alignment and asymmetry between the input and generated data. It has been noted that CycleGAN may replicate the data's "domain-specific deformations" \cite{21}, \cite{22}, which means that there are geometric changes (scaling, rotation, translation, and shear) between the source and target domains. DicycleGAN \cite{21} addressed the problem of domain-specific deformation. It used deformable convolutional (DC) layers along with new cycle-consistency losses. This model has the capacity to synthesize output data that is aligned with input data. In this model, ``deformation'' and ``image translation'' parameters were combined into a single network. However, the DC layers of DicycleGAN only learn relatively unchanged and specific deformations. The improved version of DicycleGAN was named TPSDicyc \cite{24}, formed on a thin-plate-spline (TPS). To learn the relative distortion between the input and output data, TPSDicyc used a segregated spatial transformation network (STN). The authors of the TPSDicyc technique used publicly available multi-sequence brain MR data and multi-modality abdominal data to evaluate their method, which gave good results for generating synthetic images that are aligned with the source data. The alignment issue was also tackled by \cite{22} using the alignment step, converting the source data in such a way that the position, scale, and viewing angle are identical to output data before synthesizing MR images from CT images. An improved end-to-end network, which is also based on CycleGAN, named EssNet, was introduced in \cite{23}. The EssNet synthesized unpaired MRI to CT images and CT splenomegaly segmentation concurrently in the absence of ground truth labels in CT. EssNet had a segmentation network part in addition to the CycleGAN network part. In EssNet, the alignment and domain-specific deformation issues were resolved through simultaneous training of the CycleGAN and segmentation networks. The model was trained on unpaired CT and MRI data and used ground truth only from MRI data. It generated synthetic CT images as well as the corresponding labels by leveraging the ground truth labeling of MRI images. Zhang et al. \cite{29} proposed a variant of EssNet where they used two segmentors instead of one, utilizing both source and target labels for cross-modality domain adaptation between heart MRI and CT. Similarly, Liu et al. \cite{33} also proposed a variant of EssNet where they used two segmentors instead of one, utilizing source labels only for cross-modality domain adaptation between brain and abdomen MRI and CT. More related efforts on GANs for the synthesis of medical image data for different organs are reviewed in [4], [5].

The recent state-of-the-art techniques discussed above use unpaired image to image domain adaptation to achieve image synthesis, where synthetic images from one modality with sufficient labeled data aid segmentation tasks for another modality. Our aim in this work is to use cross-modality synthesis for MRI data augmentation.
MRI image synthesis through the domain-to-domain translation of CT scan can increase the size of the dataset and thus improve the generalization of the AI models. In this work, we use a CycleGAN-based EssNet architecture for cross-modality translation of abdominal CT to abdominal MRI that can help improve the performance of a segmentation network. We train the EssNet using unpaired CT and MRI images and use manual labels for the CT images only. The simultaneous training of the CycleGAN and segmentation network parts of EssNet solves the alignment and domain-specific deformation issues. To evaluate the improvements resulting from the cross-modality translation, we use the U-Net biomedical segmentation model for liver segmentation using the original and the generated MRI images. Through the segmentation performance in terms of dice coefficient and Intersection over Union (IoU), we demonstrate the improvements that occur as a result of using the generated data. To the best of our knowledge, this is the first method that performs cross-modality abdominal (CT to MRI) image synthesis to improve the segmentation accuracy of a model. The main contributions of this work are:

\begin{itemize}

    \item We present a new two-stage EssNet+U-Net architecture that can synthesize abdominal MRI using image transformation of unpaired CT images and perform liver segmentation.
    \item We perform a cross-modality image transformation and generate synthetic MRI images that are free from alignment and domain-specific deformation issues.
    \item Through empirical evaluation, we demonstrate the potential of the synthetic MRI images and their effectiveness in improving the performance of U-Net for liver segmentation. 

\end{itemize}
The remainder of the paper is organized as follows: Section \ref{sec:methods} presents the proposed methodology in detail, followed by a discussion on datasets in Section \ref{sec:datasets}. Section \ref{sec:results} presents the results obtained with the proposed method and also presents results for ablation studies on the proposed method. Finally, Section \ref{sec:conclusion} concludes the paper.

\section{Methods}\label{sec:methods}

\begin{figure}[ht]
\centering
\includegraphics[width=0.8\textwidth]{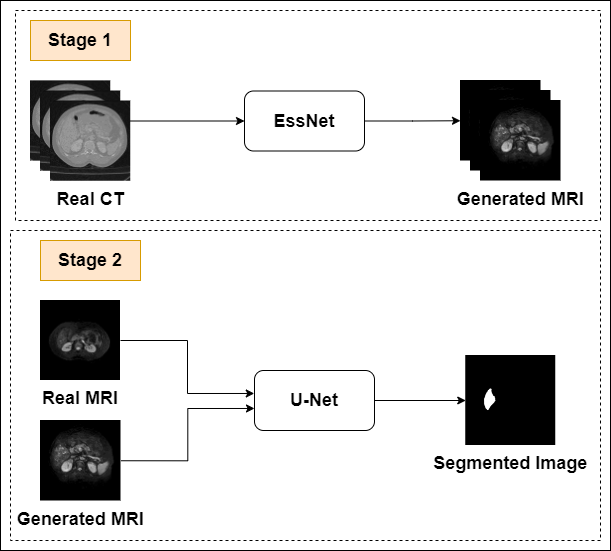}
\caption{Our proposed two-stage (EssNet+U-Net) approach. Stage 1: The approach uses CycleGAN-based EssNet architecture to generate abdominal MRI. Stage 2: The generated MRI is combined with real MRI to improve the U-Net segmentation.}
\label{fig:fig1}
\end{figure}

Huo et al., \cite{23} proposed an end-to-end synthesis and segmentation network, which performs unpaired MRI to CT image synthesis and does CT splenomegaly segmentation concurrently in the absence of ground truth labels in CT. Inspired by EssNet, we propose a two-stage (EssNet+U-Net) architecture, as shown in Figure \ref{fig:fig1}. In the proposed approach, we use EssNet to transform unpaired abdominal CT images to MRI images and perform MRI liver segmentation concurrently in the absence of ground truth labels for the MRI. The purpose of the MRI liver segmentation is to help the generator network generate high-quality MRI images without  alignment and domain-specific deformation issues. The images generated by EssNet, along with the real MRI images, are fed into the U-Net model to improve the segmentation accuracy. Unlike EssNet, which performed unpaired MRI to CT image synthesis and segmented CT splenomegaly, our aim is to use EssNet to achieve unpaired abdominal CT to MRI aligned image synthesis only and then use these newly generated images to improve the performance of U-Net segmentation. In the following text, we first describe the EssNet network details, followed by the U-Net.

\subsection{EssNet}

\begin{figure}[ht]
\centering
\includegraphics[width=0.9\linewidth]{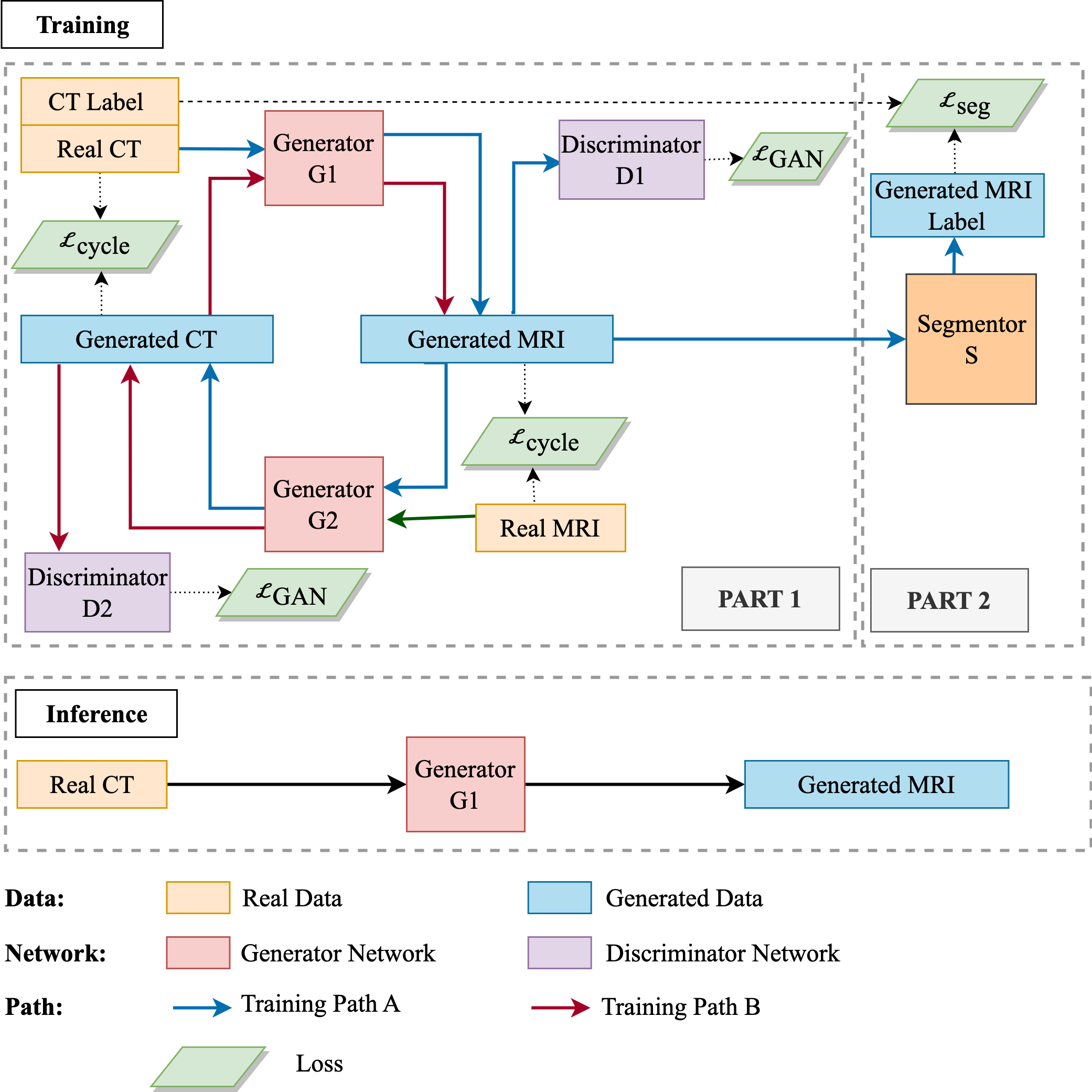}
\caption{Overall workflow of the EssNet architecture. Part 1 of the training module is the synthesis part, which is basically CycleGAN. $G_1$ and $G_2$ are the two generators, while $D_1$ and $D_2$ are the two discriminators. The segmentation part for end-to-end end training is in Part 2. In the testing module, real CT is fed to generate additional MRI images. (adapted from Huo et al. \cite{23}).}
\label{fig:fig2}
\end{figure}

Figure \ref{fig:fig2} shows the EssNet architecture. The EssNet model consists of a CycleGAN part, which has two generators and two discriminators, and a segmentation network part. Part 1 of Figure \ref{fig:fig2} is basically CycleGAN which generates synthetic images, while Part 2 of Figure \ref{fig:fig2} segments the synthetic images, which also helps to overcome the alignment problem of generated images by CycleGAN. Two generators ($G_1$ and $G_2$) are used to generate CT to MRI images and vice versa. The two adversarial discriminators ($D_1$ and $D_2$) are used for checking the probability of the input image being fake (generated) or real. $D_1$ assesses whether the MRI image is real or fake, whereas $D_2$ determines the status of the CT image. Two training routes (Path A and Path B) are present in forward cycles when this framework is applied to the unpaired CT and MRI. In the testing phase, we feed real CT images to the trained network to generate additional MRI images. The complete architecture of EssNet is discussed below, where we first discuss the generator/segmentor, followed by the discriminator, and then the loss functions.

\begin{figure}[ht!]
\centering
\includegraphics[width=0.9\linewidth]{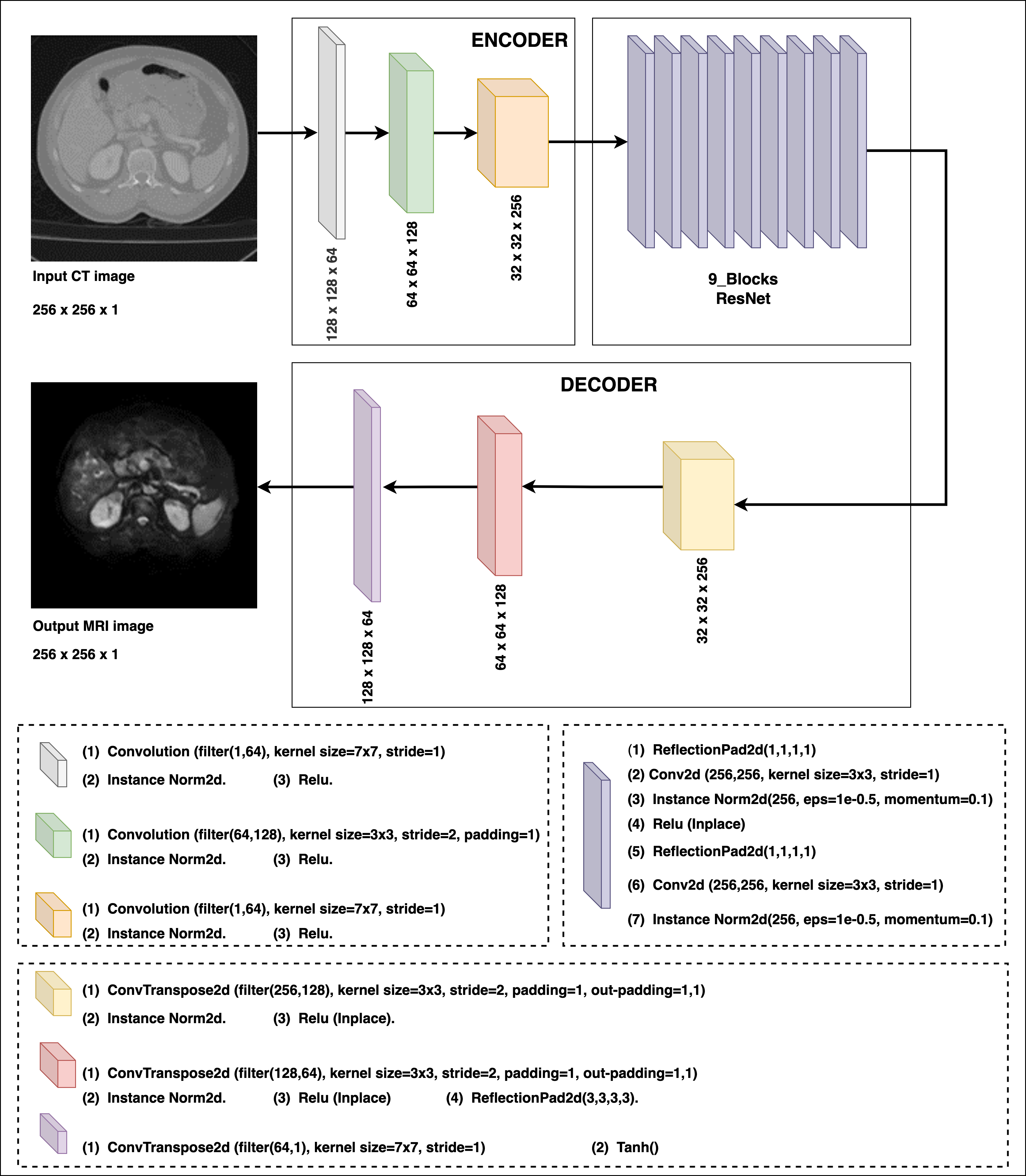}
\caption{The generator architecture includes an encoder block, a 9-block ResNet, and a decoder block.}
\label{fig:fig3}
\end{figure}

\begin{figure}[ht!]
\centering
\includegraphics[width=0.9\linewidth]{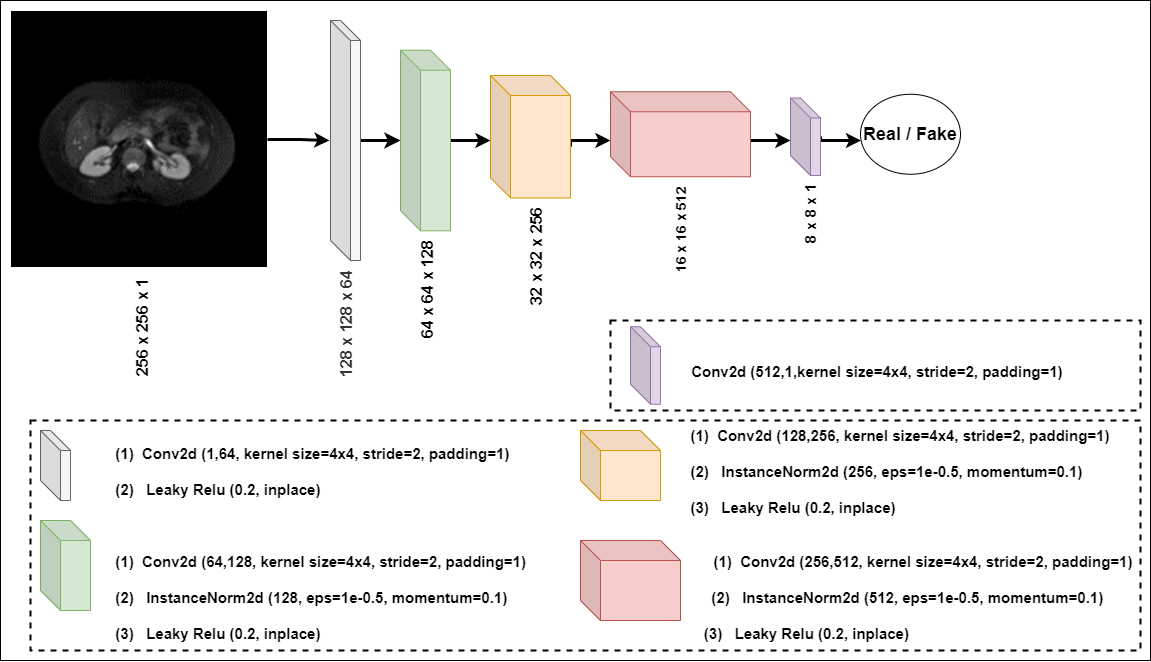}
\caption{Architecture of a patch-based discriminator. The discriminator has five convolution blocks.}
\label{fig:fig4}
\end{figure}

\subsubsection{Architecture of Generator}
The two generators ($G_1$ and $G_2$) are constructed with the 9-block ResNet (defined in \cite{17}, \cite{25}), and have an encoder-decoder architecture, as shown in Figure 3. $G_1$ transforms CT images to MRI, whereas $G_2$ transforms MRI images to CT. The segmentation network S is concatenated with $G_1$ as an additional forward branch. The 9-block ResNet \cite{17}, \cite{25} is employed as S, and its network structure is identical to that of $G_1$ and $G_2$. The segmentation is then performed on the generated MRI. ResNet employs the idea of residual blocks, which incorporate shortcut skip connections. By redefining the layers as residual functions that learn about the inputs to the layer, the authors of ResNet \cite{26} demonstrate that residual networks are simpler to optimize and can gain more accuracy at a noticeably higher depth.

\subsubsection{Architecture of Discriminator}
The architecture of two adversarial discriminators ($D_1$ and $D_2$) is patch-based, as introduced in \cite{12} and shown in Figure \ref{fig:fig4}. $D_1$ assesses whether the MRI image is real or fake, whereas $D_2$ determines the nature of the CT image. In the patch discriminator, after feeding one input image to the network, it provides us with the probabilities of two outcomes: real or fake. However, it does not utilize a scalar output; instead, it uses an N$\times$N output matrix. The output of the discriminator is an N$\times$N matrix corresponding to the N$\times$N patches of the input image. Each patch is 70$\times$70 pixels, and this size is fixed. Here, N$\times$N can vary based on the dimension of an input image.

\subsection{Loss function}
A combination of five loss functions is used to accomplish the training. The losses include two adversarial loss functions, two cycle-consistency loss functions, and one segmentation loss function, as discussed below.

\textbf{Adversarial Loss Function:}
The two adversarial loss functions, referred to as $\mathcal{L}_{GAN}(\cdot)$, are defined as follows:
\begin{align}
 \mathcal{L}_{GAN}\left(G_1, D_1, \mathrm{~A}, \mathrm{~B}\right) = E_{y \sim B}\left[\log D_1(y)\right]+E_{x \sim A}\left[\log \left(1-D_1\left(G_1(x)\right)\right)\right]
\end{align}

\begin{align}
 \mathcal{L}_{GAN}\left(G_2, D_2, \mathrm{~B}, \mathrm{~A}\right) = E_{x \sim A}\left[\log D_2(x)\right]+E_{y \sim B}\left[\log \left(1-D_2\left(G_2(y)\right)\right)\right]
\end{align}
Equations 1 and 2 are standard adversarial cross entropy loss GAN equations.
In equation 1,  $G_1$ 
the input image x of modality A, and attempts to generate images $G_1(x)$ that resembles images y of modality B, while $D_1$ attempts to differentiate between generated samples $G_1(x)$ and real images y from modality B. Equation 2 does the same thing with the roles of modality A and B reversed. The probabilities that the data instances y of modality B and x of modality A, in our case an MRI image and a CT image, are real, are estimated by the two discriminators $D_1(y)$ and $D_2(x)$. For input instances x and y, the two generators' outputs are $G_1(x)$ and $G_2(y)$. $1-D_1(G_1(x))$ and $1-D_2(G_2(y))$ estimate the probability that the input instances are synthetic. Both $E_{y\sim B}$ and $E_{x\sim A}$ represent the values that should be expected across all the real data instances. The generator tries to reduce the given loss function, whereas the discriminator tries to maximize it, similar to a min-max game.

\textbf{Cycle consistency loss function:}
Adversarial loss alone is insufficient for producing high-quality images, but the cycle consistency loss solves this problem. Cycle consistency loss is used for unpaired image-to-image translation in generative adversarial networks. Two cycle-consistency loss functions, referred to as $\mathcal{L}_{cycle}(\cdot)$, are utilized to compare the real images with the reconstructed images.
\begin{align} 
\mathcal{L}_{\text {cycle }}\left(\mathrm{G}_1, \mathrm{G}_2, \mathrm{~A}\right)=E_{x \sim A}\left[\| G_2\left(G_1(\mathrm{x})-\mathrm{x}\right)\|_1]\right. 
\end{align}
\begin{align}
\mathcal{L}_{\text {cycle }}\left(\mathrm{G}_2, \mathrm{G}_1, \mathrm{~B}\right)=E_{y \sim B}\left[\| G_1\left(G_2(\mathrm{y})-\mathrm{y} \right)\|_1]\right.
\end{align}
From equations (3) and (4):
we wish to learn a mapping between $G_1 (x)$ and $G_2 (y)$ for two domains $x$ and $y$. We wish to reinforce the idea that these translations should be inverses of each other and bijections. Therefore, the inclusion of both the cycle consistency losses for $G_2 (G_1 (x))\approx x$ and $G_1 (G_2 (y))\approx y$ are encouraged. In other words, if we pass synthesized image $G_1(x)$ (input modality A, output modality B) to $G_2$ (input modality B, output modality A), the original image x should be reconstructed and vice versa. By requiring forward and backward consistency, the loss narrows the range of potential mapping functions.

\textbf{Segmentation Loss Function:}
A segmentation loss function, referred to as $\mathcal{L}_{seg}(\cdot)$, is defined as:

\begin{align}
\mathcal{L}_{seg}(S, G_1, A) = \sum_i {N_i} \cdot \log(S(G_1(x_i)))
\end{align}

The equation is the cross entropy segmentation loss, where $S(G_1(x_i)$ shows the segmented output of segmentor S, which segments the modality B image generated by generator $G_1$, N is pixel-wise ground truth of modality A for image x, and i is the pixel index. The total loss function is subsequently defined below where the lambdas are hyper parameters set empirically:
\begin{equation}
\begin{aligned}
\mathcal{L}_{total} = \lambda_1 \cdot \mathcal{L}_{GAN}(G_1, D_1, A, B) + \lambda_2 \cdot \mathcal{L}_{GAN}(G_2, D_2, B, A) + \lambda_3 \cdot \mathcal{L}_{cycle}(G_1, G_2, A) \\ + \lambda_4 \cdot \mathcal{L}_{cycle}(G_2, G_1, B) + \lambda_5 \cdot \mathcal{L}_{seg}(S, G_1, A) 
\end{aligned}
\end{equation}

\subsection{U-Net Architecture}
We used the U-Net \cite{27} architecture to test improvements in segmentation accuracy using artificially generated images. U-Net consists of an expanding path and a contracting path. The contracting route follows the convolutional network architectural conventions. On the contracting path repeatedly, two 3x3 convolutions followed by a rectified linear unit (ReLU) are used for feature extraction, and a 2x2 max pooling operation with stride 2 is used for downsampling. With each downsampling step, we increase the number of feature channels by a factor of two. At each stage of the expanding path, the process begins with upsampling the feature map, followed by a 2x2 up-convolution. This is then concatenated with the feature map from the contracting path, which has been cropped to match in size. Afterward, two 3x3 convolutional layers are applied sequentially, each followed by a ReLU activation function. Throughout the expanding path, the number of feature channels is progressively halved at each step, which contrasts with the contracting path, where the number of channels is increased. Finally, at the last layer, each of the 64-component feature vectors is mapped to the desired number of output classes using a 1x1 convolution. The overall architecture of the network consists of 23 convolutional layers in total.

\section{Datasets}\label{sec:datasets}
This study uses two datasets: abdominal MRI (T1 and T2 weighted) and abdominal CT. The two datasets are included in the Combined (CT-MR) Healthy Abdominal Organ Segmentation (CHAOS) dataset \cite{28}. Both datasets include the MRI/CT of multiple healthy persons. The datasets were retrospectively and arbitrarily gathered from Dokuz Eylul University (DEU) Hospital's picture archiving and communication system (PACS). The CT and MR datasets are unrelated to one another since they were gathered from unregistered individuals who are distinct from one another \cite{28}.
\subsection{Abdominal CT Dataset}
\begin{figure}[h]
\centering
\includegraphics[width=0.9\linewidth]{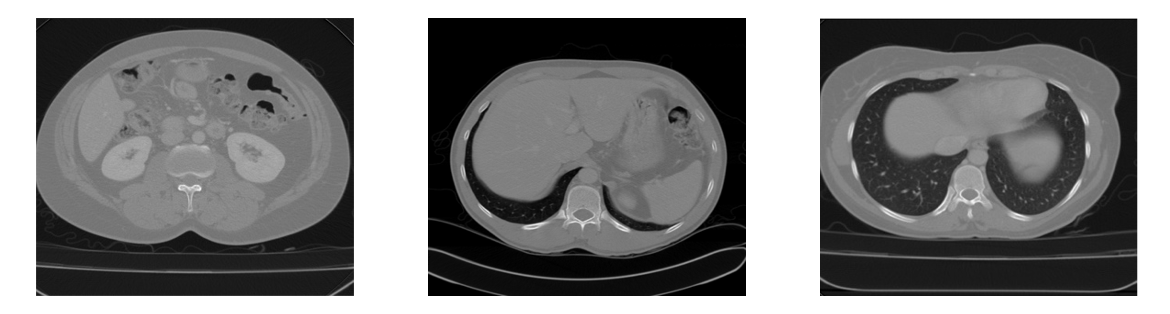}
\caption{Samples of abdominal CT images. These are the original CT images from the abdominal CT dataset and show different organs of the abdomen.}
\label{fig:fig5}
\end{figure}

There are 40 different persons' CT images in this dataset. The images in the datasets are acquired from healthy individuals, so the liver is free of lesions or other disorders. The CT images were taken from the person's upper abdomen. Each person's 3D CT DICOM file includes a collection of 2D DICOM images representing different CT cross-sections. Each DICOM image is 16-bit with a resolution of 512$\times$512. A few samples from this dataset are shown in Figure \ref{fig:fig5}. The data of 20 persons is kept for the training set, and the data of the remaining 20 persons is kept for the test set. In our work, we used only the training set for both training and testing because of the availability of ground truth CT images in this training set. The total number of images in this training set is 2874. We used those 2050 images in which the liver can be seen, as we want to do liver segmentation.
\begin{figure}[h]
\centering
\includegraphics[width=0.9\linewidth]{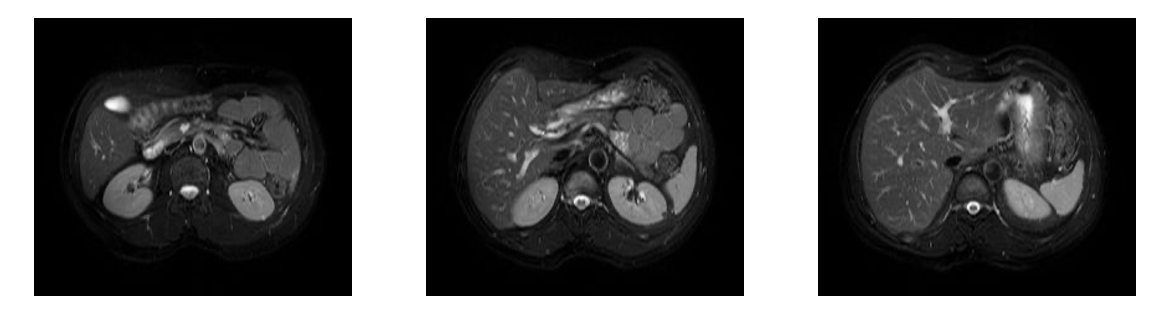}
\caption{Samples of abdominal MRI images from the T2-SPIR sequence. These are the original MRI images from the abdominal MR dataset and show different organs of the abdomen.}
\label{fig:fig6}
\end{figure}
\subsection{Abdominal MR Dataset}
In this dataset, 120 3D DICOM files from two distinct MRI sequences are included. The MRI sequences are T1-DUAL in phase (40 3D DICOM files), out phase (40 3D DICOM files), and T2-SPIR (40 3D DICOM files). Each person's 3D DICOM file includes a collection of 2D DICOM images representing different MRI cross-sections. These MRI images were also taken from the person's abdomen. Each image in this dataset is 12-bit with a resolution of 256$\times$56. A few samples from this dataset are shown in Figure \ref{fig:fig6}. From this dataset, we use T2-SPIR, which has 20 files for training and 20 files for testing. However, the ground truth is available for the training set images only, so we use samples from the training set only. The total number of images in this training set is 623. Of these, the number of images suitable for our tasks is 408, as the liver is visible in these images.
\begin{table}[h]
\centering
\caption{Number of training and test images as input to EssNet, and the number of output images by EssNet.}\label{tab:table1}
\begin{tabular}{llll}
\toprule
Train / Test & MRI & CT   & Generated MRI Images  \\
\midrule
Train        & 350 & 350  & --    \\
Test         & --   & 2050 & 2050  \\
\botrule
\end{tabular}
\end{table}

\begin{figure}[htb]
\centering
\includegraphics[width=0.4\linewidth]{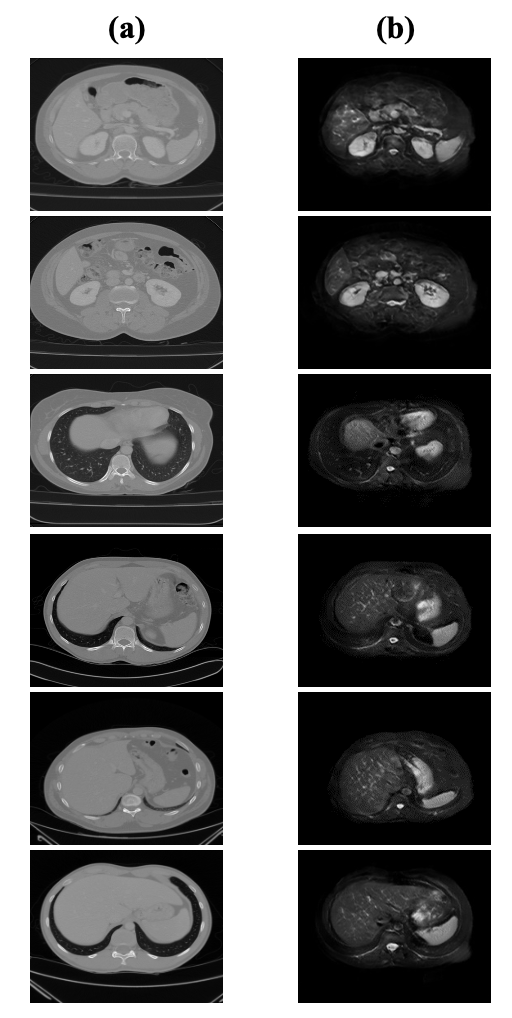}
\caption{Sample results for CT to MRI translation using EssNet on CHAOS dataset. Column (a) has the real CT images, whereas Column (b) has the generated MRI images.}
\label{fig:fig7}
\end{figure}
\begin{figure}[hb!]
\centering
\includegraphics[width=0.6\linewidth]{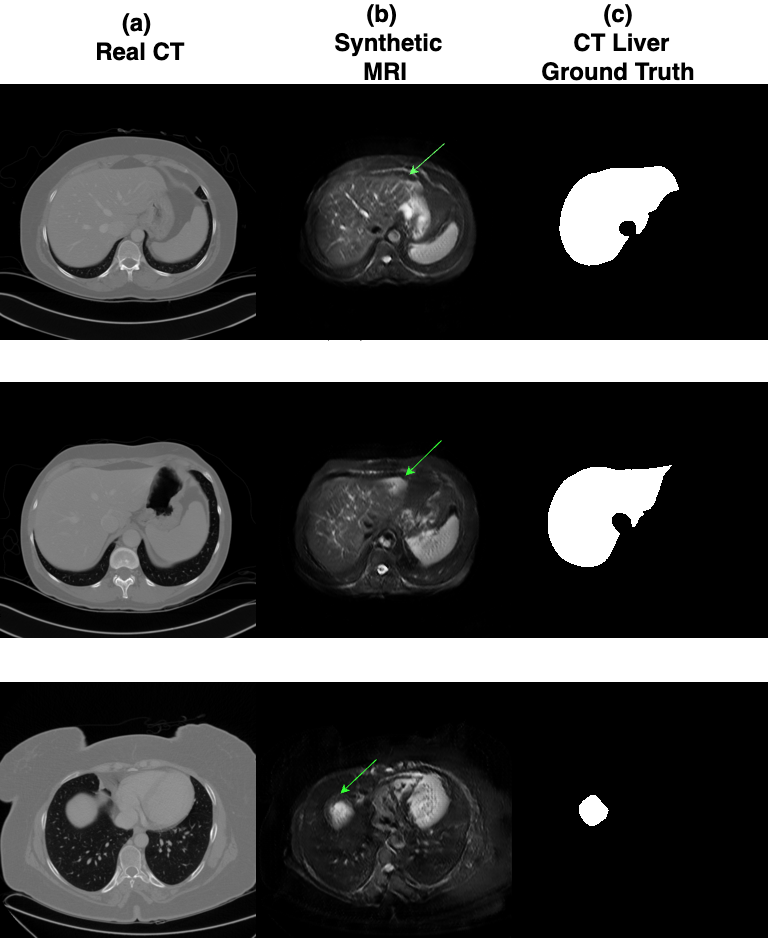}
\caption{Defects in CT to MRI translation as pointed to by arrows. Column (a) has the real CT images, Column (b) has the generated MRI images, whereas Column (c) has the CT Liver Ground Truth Segmentation Mask.}
\label{fig:fignew}
\end{figure}
\begin{figure}[h]
\centering
\includegraphics[width=0.8\linewidth]{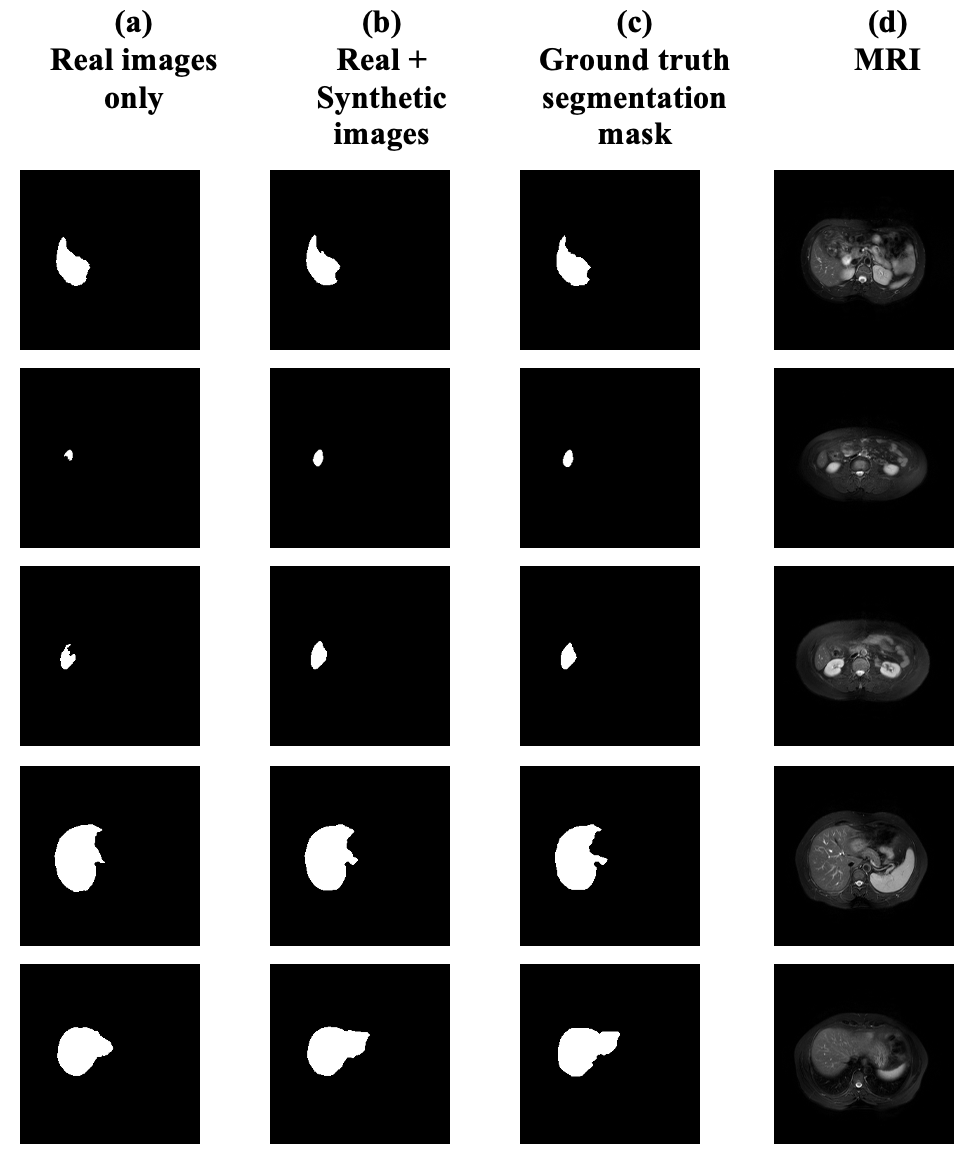}
\caption{Liver segmentation results using real and synthetic MRI. Column (a) has the segmentation results of U-Net trained only on real images; Column (b) has the segmentation results of U-Net trained on both real and synthetic images; Column (c) has the original ground truth, whereas Column (d) has the input MRI images. A close resemblance with the ground truth can be seen for segmentation when the model is trained using real as well as synthetic data.}
\label{fig:fig8}
\end{figure}
\begin{figure}[h]
\centering
\includegraphics[width=0.6\linewidth]{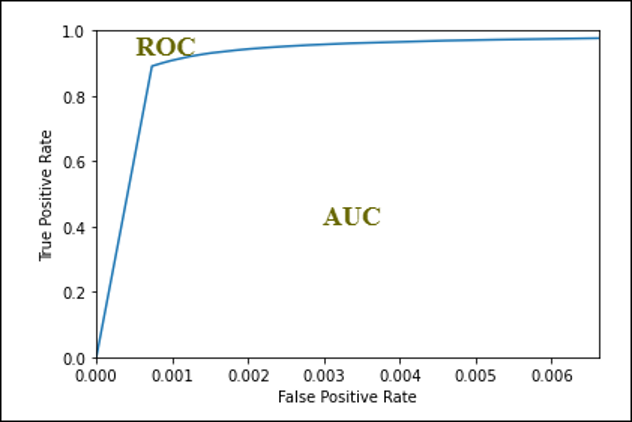}
\caption{The AUC-ROC curve. The high AUC shows that the separability between the two classes is very high.}
\label{fig:fig9}
\end{figure}
\begin{figure}[h]
\centering
\includegraphics[width=0.6\linewidth]{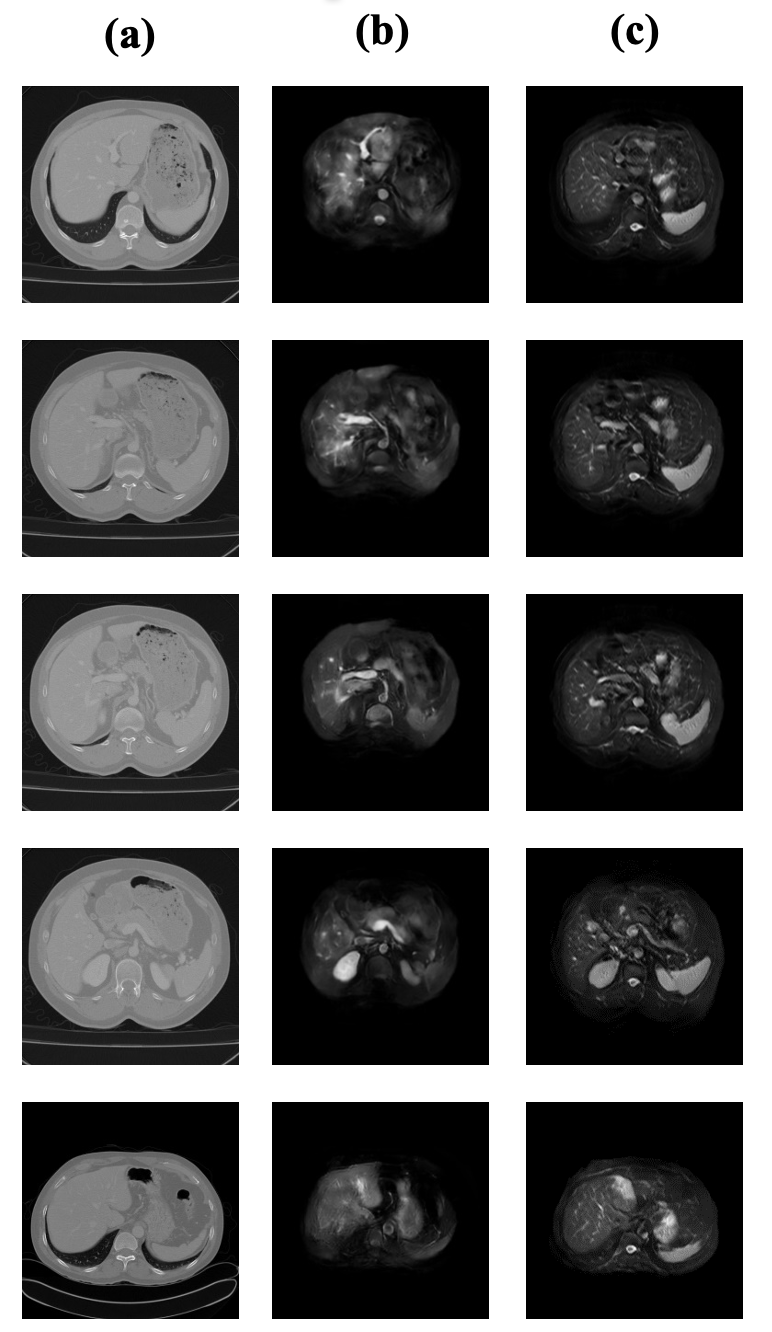}
\caption{Synthesis results for ablation studies after removing the segmentation part of the EssNet. Column (a) has the real CT images. Column (b) has the MRI images generated by CycleGAN without the segmentation part. Column (c) has the synthetic MRI images generated by EssNet architecture.}
\label{fig:fig10}
\end{figure}
\begin{figure}[h]
\centering
\includegraphics[width=\linewidth]{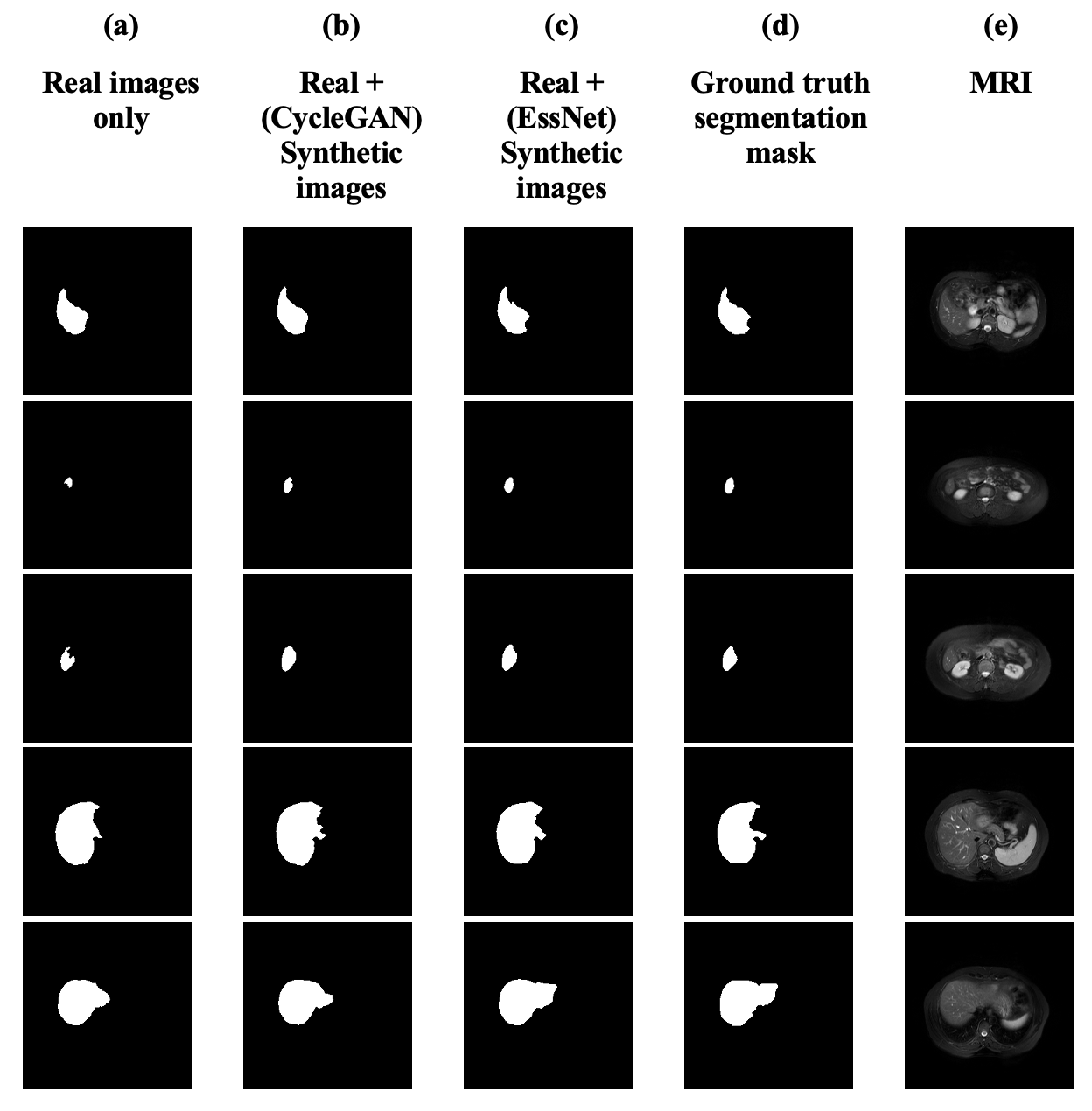}
\caption{Segmentation Results for ablation studies on MRI dataset. Column (a) has the output segmentation results of U-Net only trained on Real images; Column (b) has the output segmentation results of U-Net trained on both Real and generated images (CycleGAN), Column (c) has the output segmentation results of U-Net trained on both Real and generated images (EssNet), Column (d) has the original ground truth, whereas Column (e) has the original input MRI images. Results in (b) do not show a close resemblance to the ground truth for segmentation when the model is trained using real as well as synthesis data (CycleGAN).}
\label{fig:fig11}
\end{figure}
\begin{table}[h]
\centering
\caption{Arrangements for Training and Testing Images for U-Net.}
\label{tab:table2}
\begin{tabular}{llll}
\toprule
Train/Test & Real MRI Images & Generated MRI Images & Combined MRI Images (Real + Generated) \\
\midrule
\multirow{6}{*}{Training} & 350	& -- &	350 \\
& 350 & 354     & 704  \\
& 350 & 714     & 1064 \\
& 350 & 1034    & 1384 \\
& 350 & 1487    & 1837 \\
& 350 & 2050    & 2400 \\ \hline
Testing & 58  & -- & 58  \\
\botrule   
\end{tabular}
\end{table}

\section{Results}\label{sec:results}
The experiments were performed on an NVIDIA GeForce RTX 2060 with 6GB GDDR6 memory. For EssNet, the authors' online implementation based on PyTorch was adapted, where the loss function lambdas were set to $\lambda_1 = 1, \lambda_2 = 1, \lambda_3 = 10, \lambda_4 = 10$ based on the CycleGAN implementation whereas $\lambda_5$ was simply set to 1. The optimizer used was Adam with the learning rate of 0.0001 for $G_1$, $G_2$, and S, and 0.0002 for $D_1$ and $D_2$. The number of input and output channels for all networks was one, except for the Segmentor S. For S, we used two output channels: one for the background and one for the liver. The model was trained for 100 epochs with a batch size of 1 to remain within the GPU's memory limits. For U-Net, the Keras-TensorFlow implementation was employed. The Adam optimizer was used with the learning rate of 0.0001. The model was trained for 300 epochs with a batch size of 2, and the steps per epoch were equal to the number of images divided by the batch size.

In the first stage, we trained the EssNet network on the original 350 MRI and 350 CT images. We generated 2050 synthetic MRI images from the original CT images. The EssNet network is used only for generating new images from the original images. The arrangement of training and testing images for EssNet is shown in Table \ref{tab:table1}. Figure \ref{fig:fig7} shows the original input CT images along with their ground truth and the generated output MRI images of EssNet. It can be seen that there are no alignment or domain-specific deformation issues in the generation. However, as Figure \ref{fig:fignew} shows, there are some minor defects in the synthesized images. In each row, the first image shows the original CT image, the second image shows the synthesized MRI image, and the third shows the segmentation ground truth for the liver in the CT image. As indicated in the image by the arrow, a minor part of the liver in the synthesized image is missing, which can affect segmentation accuracy when these images are used. Therefore, in the future, more accurate synthesis of images is required, which will improve the accuracy of deep learning tasks on medical images, especially in cases where data is scarce. In the second stage, we used the U-Net architecture for liver segmentation. First, we trained the U-Net on 350 real MRI images. Then, we tested the U-Net on 58 real MRI images and calculated the dice coefficient and IoU. After this, we trained the U-Net again on real and synthetic MRI images while gradually increasing the number of synthetic images in the training data, as shown in Table \ref{tab:table2}. Finally, we tested the segmentation performance of the U-Net on 58 real MRI images. For performance evaluation, we computed the dice coefficient and IoU. The results reported in Table \ref{tab:table3} show that the dice coefficient and IoU are improving until the combined 1064 scenario. After this, the dice coefficient and IoU decrease. 
Therefore, it can be concluded that performance is saturated for U-Net when we used combined 1064 images. From Figure \ref{fig:fig8}, we can see that when U-Net segmentation models are trained using both real and synthetic data, the segmentation results show a close resemblance to the ground truth, as compared to when the U-Net is only trained on real images. The AUC-ROC curve is presented in Figure \ref{fig:fig9}, which shows the separability between the two classes. From Figure \ref{fig:fig9}, we can see that the AUC is very high, which means that the prediction capability of the model is very good.

The number of parameters for the EssNet model in our setting is 28,256,644 (28 million). The training time for 100 epochs is 6 hours and 23 minutes, i.e., 230 seconds per epoch. The test time for generating 714 images, as synthesized 714 plus real 350 gives us 1064 training images, is 352 seconds, i.e., 2.02 frames per second (FPS). The number of parameters for the U-Net model in our implementation is 31,031,685 (31 million). With 1064 training images, the training time for 300 epochs is 6 hours and 39 minutes, i.e., 80 seconds per epoch. The test time for segmenting 58 images is is 2.76 seconds i.e., 21 FPS, so the end result is a model that can segment images relatively quickly at test time. The parameters and times are summarized in Table \ref{tab:table5}. It should be noted that these training times are for an older GeForce RTX 2060 GPU, and the models are expected to train much faster on modern GPUs.

\begin{table}[h]
\centering
\caption{Results of segmentation with U-Net using varying sizes of training sets of MRI data. The size of the test set remains the same.}\label{tab:table3}
\begin{tabular}{lll}
\toprule
Number of training images (MRI) & Dice   & IoU    \\
\midrule
Real only (350)                 & 0.9459 & 0.8974 \\
Combined (704)                  & 0.9467 & 0.8989 \\
Combined (1064)                 & 0.9524 & 0.9091 \\
Combined (1384)                 & 0.9485 & 0.9020 \\
Combined (1837)                 & 0.9475 & 0.9002 \\
Combined (2400)                 & 0.9505 & 0.9058 \\
\botrule 
\end{tabular}
\end{table}
\begin{table}[h]
\centering
\caption{The output segmentation results of U-Net, trained on only real images and trained on both real and generated images by CycleGAN.}\label{tab:table4}
\begin{tabular}{lll}
\toprule
Number of training images (MRI) & Dice   & IoU    \\
\midrule
Real images only (350) & 0.9459 & 0.8974 \\
Combined (1064) Real+Synthetic & 0.9460 & 0.8976\\
\botrule       
\end{tabular}
\end{table}

\begin{table}[h]
\centering
\caption{Computational Cost of the Models}\label{tab:table5}
\begin{tabular}{llll}
\toprule
 & Trainable Parameters & {Train Time (seconds per epoch)} & {Inference (frames per second)}\\
\midrule
EssNet & 28 M & 230 & 2.02\\
U-Net & 31 M & 80 & 21\\
\botrule       
\end{tabular}
\end{table}

\subsection{Ablation Studies}
To illustrate the viability of the proposed model, we performed an ablation study on the synthesized MRI images from CT images. We made changes to the EssNet model by removing the segmentation network (part 2 in Figure \ref{fig:fig2}), i.e., we used the CylceGAN only. We generated MRI images from CT images through CycleGAN and kept the training and testing images the same as EssNet, as discussed in the results section. Figure \ref{fig:fig10} shows the results of CycleGAN for CT to MR synthesis. There are some alignment and domain-specific deformations in the generated images. To see the effect of this deformation, we trained U-Net using both real and synthetic MRI images and calculated the dice coefficient and IoU. The results reported in Table~\ref{tab:table4} show that using the real and synthetic MRI images did not bring any improvements in dice coefficient and IoU compared to the results for real images only. Figure \ref{fig:fig11} also shows that there is little or no difference in the output segmentation results of U-Net, trained on only real images and trained on both real and synthetic images and that the segmentation results do not show a close resemblance to the ground truth. From this ablation study, we can say that EssNet based method is better at synthesizing cross-modality images, as these images have no alignment or deformation issue.

\subsection{Limitations}
The results presented in this study are derived from a dataset originating from a single hospital, and their generalizability to other datasets/hospitals remains unverified. While the framework demonstrates promising performance within this dataset, caution is advised when extrapolating these findings. The segmentation performance presented in this work is limited to the use of the U-Net model only and does not explore other advanced methods for segmentation. We believe it is justified as the focus of this work has been to show the effectiveness of the synthetic data, and the use of more advanced segmentation techniques has been left out for future experiments. 



\section{Conclusion}\label{sec:conclusion}
Medical image synthesis using GANs is an exciting and rapidly evolving field that holds great potential for improving medical image analysis and computer-aided diagnosis. In this work, we proposed a two-stage (EssNet+U-Net) architecture for improving the accuracy of liver segmentation through cross-modality medical image synthesis. We use the EssNet model to transform unpaired abdominal CT images to MRI images in such a way that there are no alignment issues or domain-specific deformation issues. The images generated by EssNet are fed into the U-Net model along with the real images to improve segmentation. We evaluated the performance of U-Net in both cases, i.e., only trained on real images and trained on both real and synthetic images. Experimental results showed the effectiveness of our proposed method, where the segmentation accuracy results improved when we trained U-Net on real plus synthetic images. In summary, the improvement achieved is 1.17\% in IoU and 0.65\% in the dice coefficient. We believe that these results can serve as motivation for future research into generating cross-modality images in order to increase the number of images in a dataset, particularly, in clinical applications,  without the need to acquire additional data where training data is scarce and acquiring additional data is costly. The improvements of the order 1.17\% and 0.65\% in the IoU and the dice coefficient are relatively smaller; however, it is not uncommon to report such marginal improvements for segmentation tasks. Since the focus of the work has been to demonstrate the effectiveness of the synthetic image data, only U-Net was used for the segmentation task. It is expected that more advanced segmentation methods would bring further improvements; however, their use is left out for future work. 

\backmatter

\section*{Statements and Declarations}
\begin{itemize}
\item Funding: No funding was received for conducting this study.
\item Competing interests: No potential competing interest was reported by the author(s).
\item Ethics approval: Not applicable. 
\item Consent to participate: Not applicable.
\item Consent for publication: Not applicable.
\item Availability of data and materials: All data analysed during this study is publicly available from https://chaos.grand-challenge.org/.
\item Code availability: Not applicable.
\item Authors' contributions: H. A. and S. A. designed the research. M. R. conducted the experiments. M. R., H. A. and S. A. did the analysis and interpretation of the results. G. M., and S. A. supervised the work. M. R. and S. A. wrote the manuscript. H. A., G. M., and Z. S. revised the manuscript. All authors reviewed the manuscript. 
\end{itemize}

\bigskip



\end{document}